\documentstyle[epsfig]{aipproc}

\newcommand{\be}{\begin{equation}}
\newcommand{\bea}{\begin{eqnarray}}
\newcommand{\en}{\end{equation}}
\newcommand{\eea}{\end{eqnarray}}

\newcommand{\fract}[2]{{\textstyle\frac{#1}{#2}}} 
\begin{document}
\title{Radial Excitations of low--lying Baryons and the 
Structure of the Z$^+$ Penta--Quark\footnote{Talk presented 
at the $7^{\rm th}$ Conference on the Intersections of Particle 
and Nuclear Physics, Quebec, May 2000. To appear in the 
proceedings.}
}

\author{Herbert Weigel\thanks{Heisenberg--Fellow}}
\address{Center for Theoretical Physics, Laboratory for
        Nuclear Science,
        and Department of Physics \\
        Massachusetts Institute of Technology,
        Cambridge, Massachusetts 02139\\~\\
{\rm MIT-CTP\#2990}~~~{\rm hep-ph/0006191}
}

\maketitle

\begin{abstract}
Within the collective quantization scheme for chiral solitons
we discuss states in higher dimensional representations of 
flavor $SU(3)$ and their relation to radially excited states in
the octet. We also consider states which do not have
counterparts of the same quantum numbers in the octet 
or decuplet and cannot be built from three quarks. 
We focus on the $Z^+$ penta--quark, presumably
the lightest such state, by estimating its mass and decay 
width.
\end{abstract}
\section*{Introduction}
In the collective quantization of chiral solitons higher dimensional 
representations of flavor $SU(3)$ play a major role\footnote{See
ref \cite{We96} for a review and references on chiral solitons in 
flavor $SU(3)$.}. Upon flavor symmetry breaking, members of these
representations admix to the ordinary octet and decuplet states 
of the same quantum numbers to form the low--lying $\fract{1}{2}^+$ and 
$\fract{3}{2}^+$ baryons. Notably these higher dimensional 
representations also contain states with quantum numbers 
that are not found in the octet (decuplet) representation. In a quark
model picture these baryons cannot be formed as simple three 
quark states, rather they consist of at least four quarks and an 
anti--quark. The $\overline{\bf 10}$ representation contains the 
presumably lightest such state,
the $Z^+$, with the quantum numbers $Y=2$, $I=0$, $J=1/2$.
Details of the presented study are found 
in ref \cite{We98}. For earlier studies on the $Z^+$ see ref~\cite{Wa92}.
An experimental search for the $Z^+$ was proposed
in ref~\cite{Pa99}.

\section*{Collective quantization}

We consider a chiral Lagrangian in flavor $SU(3)$. The basic
variable is the chiral field $U={\rm exp}(i\lambda_a\phi^a/2)$
representing the pseudoscalar fields $\phi^a$ ($a=0,\ldots,8$). 
Other fields may be included as well. For example, the specific model 
used later also contains a scalar meson. In general this chiral 
Lagrangian can be written as 
\be
{\cal L}={\cal L}_{\rm S}+{\cal L}_{\rm SB}
\label{lag1}
\en
with flavor symmetric (S) and flavor symmetry breaking (SB) pieces.
Denoting the (classical) soliton solution of (\ref{lag1}) by 
$U_0(\vec{r})$ states with baryon quantum numbers are 
constructed by quantizing the flavor rotations 
\be
U(\vec{r},t)=A(t)U_0(\vec{r})A^\dagger(t), \qquad
A(t)\in SU(3)
\label{collcor1}
\en
canonically. According to (\ref{lag1}) the Hamiltonian for the collective 
coordinates $A(t)$ can be written as $H=H_{\rm S}+H_{\rm SB}\, .$
For unit baryon number the eigenstates of $H_{\rm S}$ are the members 
of $SU(3)$ representations with the condition that the representation
contains a state with identical spin and isospin quantum numbers
(such as {\it e.g.} the nucleon or the $\Delta$). The symmetry breaking 
piece, $H_{\rm SB}$ mixes states from different $SU(3)$ representations.
For example, the state identified with the nucleon becomes
\be
|N\rangle=|N,{\bf 8}\rangle+
c^{(N)}_{\overline{\bf 10}}|N,\overline{\bf 10}\rangle+
c^{(N)}_{\bf 27}|N,{\bf 27}\rangle+\ldots\,\, ,
\label{mix1}
\en
where the coefficients are computed as matrix elements of $H_{\rm SB}$.
Another interesting question, of course, is what role do states play
which diagonalize $H$ and have dominant contributions from the 
higher dimensional representations such as
\bea
|N^\prime\rangle&=&|N,\overline{\bf 10}\rangle+
c^{(N^\prime)}_{\bf 8}|N,{\bf 8}\rangle+
c^{(N^\prime)}_{\bf 27}|N,{\bf 27}\rangle+\ldots\,\, ,
\label{mix2}\\
|Z^+\rangle&=&|Z^+,\overline{\bf 10}\rangle+
c^{(Z^+)}_{\overline{\bf 35}}|Z^+,\overline{\bf 35}\rangle
+\ldots\,\, ?
\label{mix3}
\eea
A na\"{\i}ve analysis predicts the $N^\prime$ to be 
$400-500{\rm MeV}$ heavier than the nucleon which suggests to identify
this state with the Roper (1440) resonance. However, such an 
interpretation contradicts the common folklore viewing the Roper 
as a radial 
excitation of the nucleon. In order to resolve this puzzle and also
to provide a sensible interpretation to the state $|Z^+\rangle$,
the radial degrees of freedom need to be quantized as well. For this
end we introduce the corresponding collective coordinate $\xi(t)$ 
via~\cite{Sch91a,Sch91b}
\be
U(\vec{r},t)=A(t)U_0(\xi(t)\vec{r})A^\dagger(t)\, .
\label{collcor2}
\en
Changing variables to $x(t)=[\xi(t)]^{-3/2}$ the flavor symmetric
piece of the collective Hamiltonian can be diagonalized for
a given $SU(3)$ representation of dimension $\mu$
\be
H_{\rm S}=\frac{-1}{2\sqrt{m\alpha^3\beta^4}}\frac{\partial}{\partial x}
\sqrt{\frac{\alpha^3\beta^4}{m}}\frac{\partial}{\partial x}
+V+\left(\frac{1}{2\alpha}-\frac{1}{2\beta}\right)J(J+1)
+\frac{1}{2\beta}C_2(\mu)+s\, ,
\label{freebreath}
\en
where $J$ and $C_2(\mu)$ are the spin and (quadratic) Casimir
eigenvalues associated with the representation $\mu$. Note that
$m=m(x),\alpha=\alpha(x),\ldots,s=s(x)$ are all functions of the 
scaling variable.
We use (\ref{freebreath}) to build towers of radially excited
states $|\mu,n_\mu\rangle$ with $n_\mu=0,1,\ldots$ denoting the number 
of nodes in the respective wave--functions. These states serve
to diagonalize the Hamiltonian including symmetry breaking
\be
H_{\mu,n_\mu;\mu^\prime,n^\prime_{\mu^\prime}}=
{\cal E}_{\mu,n_\mu}\delta_{\mu,\mu^\prime}
\delta_{n_\mu,n^\prime_{\mu^\prime}}
-\langle\mu,n_\mu 
|\fract{1}{2}{\rm tr}\left(\lambda_8A\lambda_8A^\dagger\right)
s(x)|\mu^\prime,n^\prime_{\mu^\prime}\rangle \ ,
\label{hammatr}
\en
yielding the baryonic states
$|B,m\rangle=\sum_{\mu,n_\mu}C_{\mu,n_\mu}^{(B,m)}
|\mu,n_\mu\rangle \,.$
The resulting spectrum~\cite{Sch91b} is depicted in figure~1 
and compared to the known resonances \cite{PDG98}. We recognize 
a good agreement although the Roper is predicted about $90{\rm MeV}$
too low. It also worth to note that this state turns out to 
dominantly be a radial excitation of the nucleon state in the 
octet representation, {\it i.e.}~$|{\bf 8},1\rangle$~\cite{We98}.

~\\
\parbox[l]{10.1cm}{
\centerline{
\setlength{\unitlength}{0.5mm}
\begin{picture}(20,120)
\thicklines
\put(0,0){\line(1,0){160}}
\put(0,0){\line(0,1){115}}
\put(0,115){\vector(0,1){5}}
\put(-1,50){\line(1,0){2}}
\put(-1,100){\line(1,0){2}}
\put(-30,118){\large {${\frac{M_B-M_N}{{\rm GeV}}}$}}
\put(-20,49){$0.5$}
\put(-20,99){$1.0$}
\put(13,-8){$N$}
\put(45,-8){$\Lambda$}
\put(77,-8){$\Sigma$}
\put(109,-8){$\Xi$}
\put(141,-8){$\Delta$}
\thinlines
\multiput(0,50)(3,0){10}{\line(1,0){1}}
\put(4,53){\footnotesize{$Roper$}}
\multiput(0,77)(3,0){10}{\line(1,0){1}}
\multiput(32,17.7)(3,0){10}{\line(1,0){1}}
\multiput(32,66.3)(3,0){10}{\line(1,0){1}}
\multiput(32,87)(3,0){10}{\line(1,0){1}}
\multiput(64,25)(3,0){10}{\line(1,0){1}}
\multiput(64,72)(3,0){10}{\line(1,0){1}}
\multiput(64,94)(3,0){10}{\line(1,0){1}}
\multiput(96,37.9)(3,0){10}{\line(1,0){1}}
\multiput(128,29)(3,0){10}{\line(1,0){1}}
\multiput(128,66)(3,0){10}{\line(1,0){1}}
\multiput(128,98)(3,0){10}{\line(1,0){1}}
\thicklines
\put(0,41){\line(1,0){28}}
\put(0,84){\line(1,0){28}}
\put(0,87){\line(1,0){28}}
\put(32,17.5){\line(1,0){28}}
\put(32,65.8){\line(1,0){28}}
\put(32,108){\line(1,0){28}}
\put(64,28){\line(1,0){28}}
\put(64,69){\line(1,0){28}}
\put(64,107){\line(1,0){28}}
\put(96,38.1){\line(1,0){28}}
\put(96,94){\line(1,0){28}}
\put(128,26){\line(1,0){28}}
\put(128,64){\line(1,0){28}}
\put(128,94){\line(1,0){28}}
\end{picture}\hspace{6.2cm}
}}\parbox[r]{4.6cm}{\tenrm 
$\scriptsize{\mbox{\boldmath ${\rm FIGURE~1:}$}}$
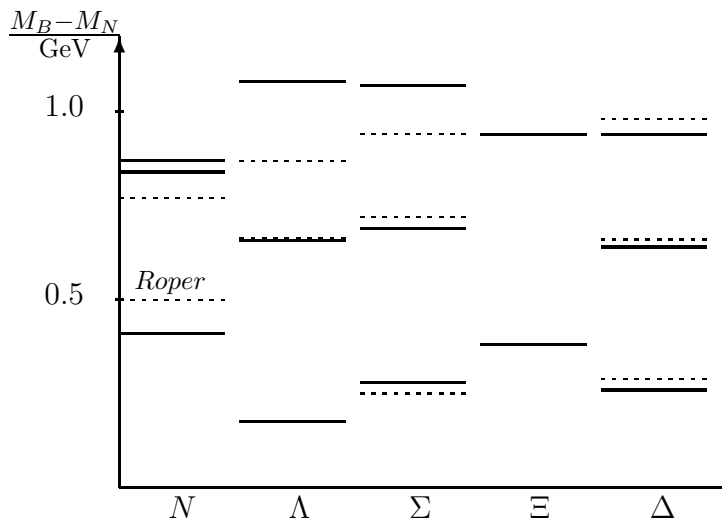
The mass differences of the predicted baryons in the scaling mode 
treatment of the three flavor Skyrme model with a scalar field
included (full lines) \protect\cite{Sch91b}; empirical data 
\protect\cite{PDG98} are denoted by dotted lines.
The model predictions for the ground states of the 
$\Lambda$ and $\Xi$ channels and the experimentally 
observed counterparts are (almost) indistinguishable.}
~\\~\\~\\
For the baryon state we are interested in most, we find
$M_{Z^+}-M_N\approx630 {\rm MeV}\,.$
This state has a sizable admixture of the counterpart
in the $\overline{\bf 35}$ representation.

\section*{Estimate of decay widths}
In order to further specify the $Z^+$ we demand the width for 
the decay $Z^+\to KN$. This is one of the strong decays of
excited baryons which are described by
\be
\Gamma_{B^\prime\to B \phi}=\frac{3G_{B^\prime\to B \phi}^2}
{8\pi M_{B^\prime}M_B} \ |\mbox{\boldmath $p$}_\phi|^3 \ .
\label{width1}
\en
Here $\mbox{\boldmath $p$}_\phi$ is the momentum of the outgoing meson
in the rest frame of~$B^\prime$. The coupling constants
$G_{B^\prime\to B \phi}$ are not easily accessible in chiral soliton
models. Fortunately, it seems sufficient to estimate them from the
appropriate flavor operator (For example, $a=3$ when $B$ and $B^\prime$
have identical isospin projections, $I_3$.),
\be
G_{B^\prime\to B \phi}= C_{\Delta}
\langle B^\prime m^\prime | x^{2n/3} 
{\rm tr}\left(\lambda_a A\lambda_3 A^\dagger\right)
 | B m \rangle 
\label{width2}
\en
and allow for different forms for the scaling operators parameterized by 
the power $n$. The constant $C_{\Delta}$ is fixed such that 
the experimental width $\Gamma_{\Delta\to\pi N}=120{\rm MeV}$
is properly reproduced. From table~\ref{tab_3}
we observe that only the width of the decay $R\to N\pi$ 
has a significant dependence on the shape of the scaling function
in (\ref{width2}). 
\begin{table}[ht]
\caption{\label{tab_3}Decay widths (in MeV) and ratio of $\pi N$ and
$\pi\Delta$ coupling constants using the matrix elements
(\protect\ref{width2}). $R$ denotes the
Roper (1440) resonance. Experimental data are extracted
from ref \protect\cite{PDG98}.}
\begin{tabular}{c | c c c | c c c| c}
& \multicolumn{3}{c|}{e=5.0} & \multicolumn{3}{c|}{e=5.5} & Data \\
\hline
$n$ & 4 & 3 & 2 & 4 & 3 & 2 & \\
\hline
$\Sigma^*\to\Sigma\pi$  & 1 & 1 & 1 & 2 & 2 & 2 & $4\pm1$ \\
$\Sigma^*\to\Lambda\pi$ & 33& 38& 42& 37& 38& 43& $32\pm4$ \\
$\Xi^* \to \Xi \pi $    & 5 & 7 & 10& 7 & 9 & 11& $10\pm2$ \\
\hline
$R\to N\pi $ & 429 & 281 & 156 & 424 & 260 & 145 & 200 to 320 \\
$R\to\Delta\pi $ & 4 & 2 & 2 & 9 & 6 & 3 & 50 to 80 \\
\hline
$Z^+\to N K$ & 118 & 121 & 124 & 130 & 124 & 126 & ? \\
\hline
$g_{\pi NN}/g_{\pi N\Delta} $ &
0.77 & 0.79 & 0.83 & 0.77 & 0.79 & 0.83 & 0.68
\end{tabular}
\end{table}
The width for $R\to\Delta\pi$ is underestimated because the respective
phase space is small as a consequence of the too low prediction 
for the Roper mass. Otherwise the widths are reasonably
well reproduced and we estimate
$\Gamma_{Z^+\to KN}\approx 120{\rm MeV}.$

\section*{Conclusion}
We have investigated the role of higher dimensional flavor $SU(3)$ 
representations for the description of excited baryon states. In particular 
the interplay between the states in these representations and the radial
excitations of the octet (and decuplet) baryons has been discussed.
A reasonable description of the spectrum and the decay widths has
been achieved. The lightest of the exotic states in such a 
higher dimensional representation, the $Z^+$, was predicted to have 
a mass of about $1.57{\rm GeV}$ and a width of about 
$120{\rm MeV}$ for its decay into a nucleon and a kaon.
\subsection*{Acknowledgements}
This work is supported in part by funds provided by
the U.S. Department of Energy (D.O.E.) under cooperative research agreement
\#DF-FC02-94ER40818 and the Deutsche Forschungsgemeinschaft (DFG) under
contract We 1254/3-1.

\end{document}